\documentclass[aps,prl,floatfix,superscriptaddress,twocolumn,tightenlines,prl,nofootinbib,10pt]{revtex4}
\usepackage{graphicx}% Include figure files
\usepackage{dcolumn}% Align table columns on decimal point
\usepackage{bm}% bold math
\usepackage{color}% add hypertext capabilities
\begin{document}
\preprint{APS/123-QED}

\title{Non-line-of-sight 3D imaging with a single-pixel camera}

\author{G. Musarra}
\email{2376730m@student.gla.ac.uk}
\affiliation{School of Physics \& Astronomy, University of Glasgow, Glasgow G12 8QQ, UK.}

\author{A. Lyons}
\affiliation{School of Physics \& Astronomy, University of Glasgow, Glasgow G12 8QQ, UK.}

\author{E. Conca}
\affiliation{Dipartimento di Elettronica, Informazione e Bioingegneria, Politecnico di Milano, 20133 Milano, Italy}

%\author{M. P. Edgar}
%\affiliation{School of Physics \& Astronomy, University of Glasgow, Glasgow G12 8QQ, UK.}

\author{Y. Altmann}
\affiliation{School of Engineering and Physical Sciences, Heriot-Watt University, Edinburgh EH144AS, UK.}

\author{F. Villa}
\affiliation{Dipartimento di Elettronica, Informazione e Bioingegneria, Politecnico di Milano, 20133 Milano, Italy}

\author{F. Zappa}
\affiliation{Dipartimento di Elettronica, Informazione e Bioingegneria, Politecnico di Milano, 20133 Milano, Italy}

\author{M. J. Padgett}
\affiliation{School of Physics \& Astronomy, University of Glasgow, Glasgow G12 8QQ, UK.}

\author{D. Faccio}
\email{Daniele.Faccio@glasgow.ac.uk}
\affiliation{School of Physics \& Astronomy, University of Glasgow, Glasgow G12 8QQ, UK.}

\date{\today}
\begin{abstract} 
Real time, high resolution 3D reconstruction of scenes hidden from the direct field of view is a challenging field of research with applications in real-life situations related e.g. to surveillance, self-driving cars and rescue missions. Most current techniques recover the 3D structure of a non-line-of-sight (NLOS) static scene by detecting the return signal from the hidden object on a scattering observation area. Here, we demonstrate the full colour retrieval of the 3D shape of a hidden scene  by coupling back-projection imaging algorithms with the high-resolution time-of-flight information provided by a single-pixel camera. By using a high efficiency Single-Photon Avalanche Diode (SPAD) detector, this technique provides the advantage of imaging with no mechanical scanning parts, with acquisition times down to sub-seconds.
	
	% \begin{description}
	% \item[PACS numbers] 
	%  \verb+05.45.-a+, \verb+42.65.Sf+, \verb+05.45.Yv+,  \verb+42.65.Hw+  
	% \end{description}
\end{abstract}
\pacs{Valid PACS appear here}
\maketitle
\begin{figure}[t!]
	\centering
	{\includegraphics[width=\linewidth]{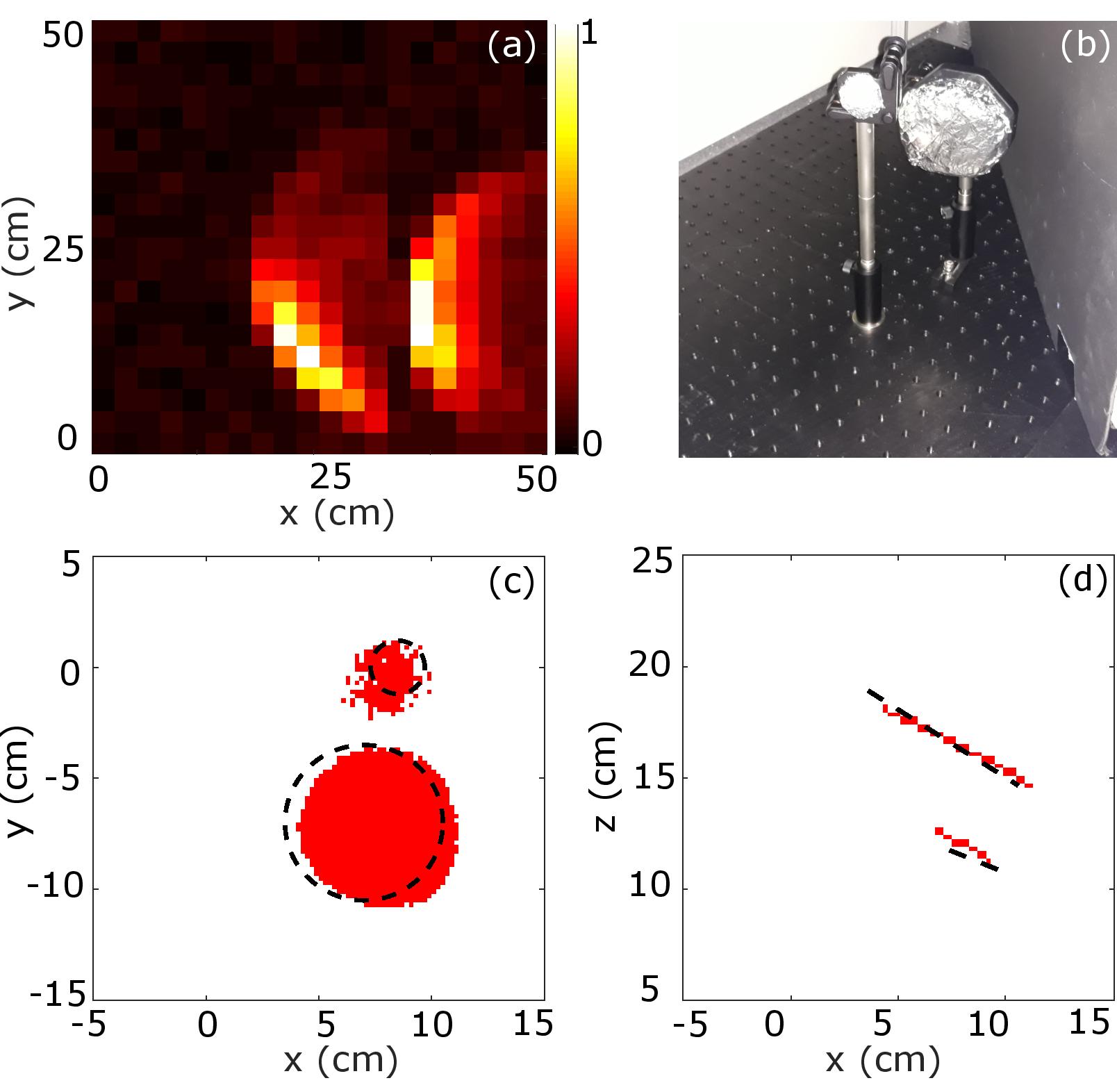}} \\
	\caption{Cooperative two objects scenario for NLOS single-pixel camera based on a PMT detector. a) Return signal  produced by the two hidden objects and retrieved by raster scan imaging onto the DMD and collecting the signal on the single-pixel detector at the time frame $t=7.1$ ns. c-d) 3D shape retrieval of the hidden scene in the x-y plane (c) and in the x-z plane (d) after applying backprojection imaging algorithm.  The dotted line represents the actual position of the targets. To facilitate the visualization  the reflectivity is normalized to the local maxima value. The objects to be recovered are two round targets of 2.54 cm and 7.62 cm of diameter at a varying depth and tilted. The voxels dimension is  0.2x0.2x1 cm$^3$ for an area  to investigate of 20x20x100 cm$^3$. The threshold used for the retrieval is 0.89.}
	\label{2}
\end{figure}
\begin{figure}[t!]
	\centering
	{\includegraphics[width=\linewidth]{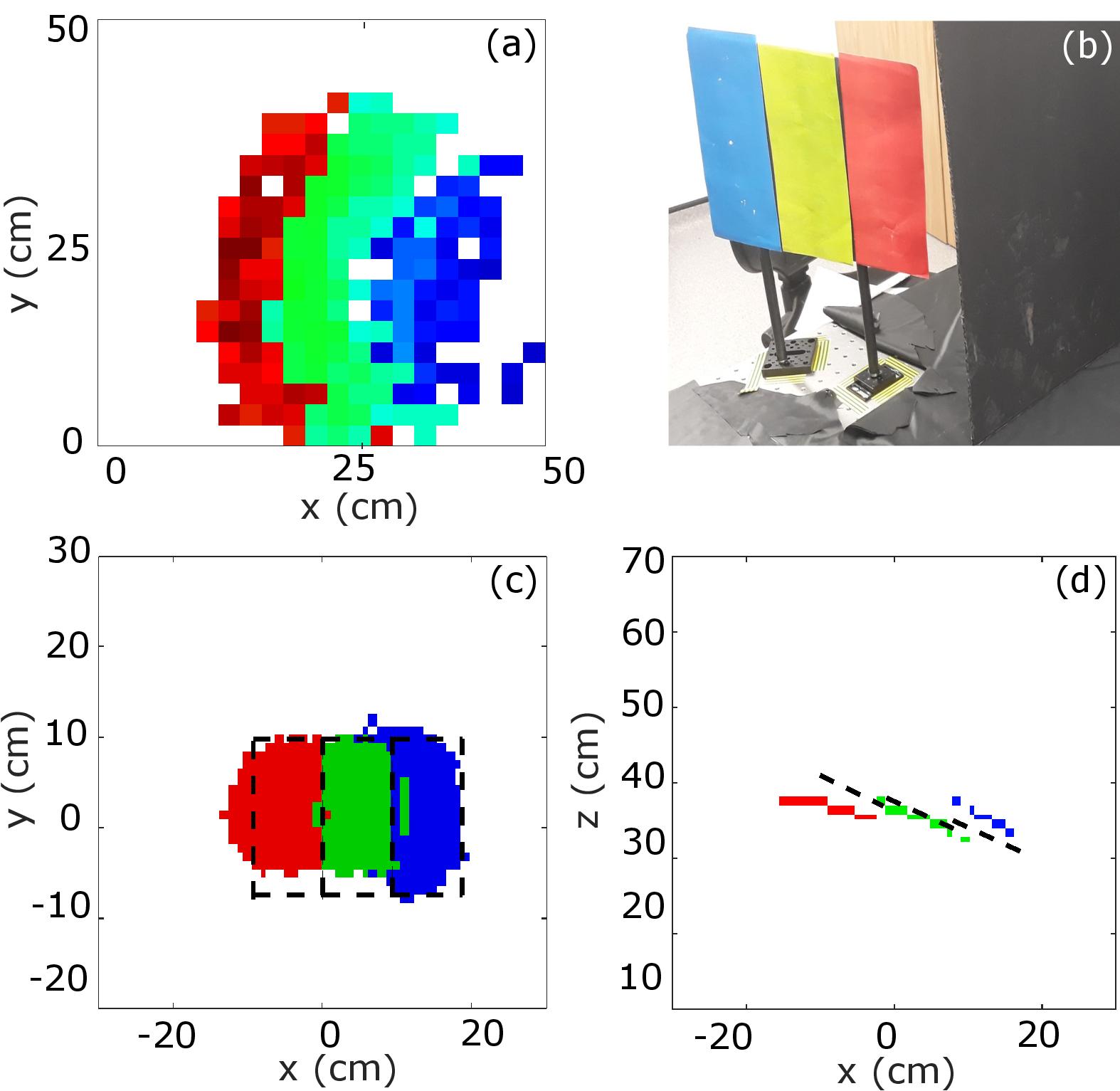}} \\
	\caption{Non cooperative objects  RGB coloured scenario for NLOS single-pixel camera  based on a PMT detector. The red colour corresponds to the recovered target by using the red spectral filter and so on.  a) Return signal  produced by the hidden object by raster scan acquisition. b) Picture of the RGB coloured target. c,d) RGB retrieval of the 3D shape of the hidden objects in the x-y plane (c) and in the x-z plane (d) after applying backprojection imaging algorithm. The red colour corresponds to the recovered target by using the red spectral filter and so on. The dotted black line represents the actual position of the target. We discretise the space in voxels of 1.4x1.4x1 cm$^3$ for an overall area of 140x140x100 cm$^3$. }%The threshold used for the retrieval is  0.90.}
	\label{3}
\end{figure}

The identification of scenes hidden from the direct line-of-sight, as happens for objects hidden behind an occluder or wall, is a challenging imaging task with applications in defence, surveillance and self-driving vehicles \cite{faccio2018trillion}.  Non-line-of-sight (NLOS) imaging has been demonstrated by using radar systems \cite{sume2009radar}, wavefront shaping \cite{katz2012looking} and speckle correlation \cite{katz2014non,faccio2018trillion,altmann2018quantum} and recently even with passive cameras capturing light originating from behind a wall using an ordinary digital camera \cite{goyal2019}. Most approaches in this field have demonstrated how to identify the hidden scene by collecting the light scattered back by  hidden objects with a system similar to light detection and ranging (LIDAR) by using the time-of-flight information of the back scattered signal \cite{wandinger2005introduction, schwarz2010lidar, kirmani2011looking, buttafava2015non,pediredla2017reconstructing}. This technique typically involves a pulsed laser beam pointed on a scattering surface, producing a spherical wave propagating into the hidden scene. When the spherical wave hits the hidden object, the light is then scattered back towards the scattering surface. Collecting the third bounce echo scattered from the hidden object allows the detection and identification of hidden scene by advanced 3D reconstruction algorithms \cite{velten2012recovering}. Past results have demonstrated how this technique can be used for tracking a moving hidden object even over large distances  \cite{gariepy2016detection,chan2017non} and for the retrieval of the 3D shape of a static hidden object by using  back-projection imaging algorithms \cite{velten2012recovering,Arellano17} or ellipsoid mode decomposition for multiple hidden objects \cite{jin2018reconstruction}.  Alternative methods  aimed at simplifying or increasing the speed of  LIDAR-like NLOS imaging rely on 2D continuous illumination \cite{klein2016tracking}, deep learning \cite{caramazza2018neural} and confocal illumination/collection \cite{o2018confocal}. The multiple-bounce back-scattered signal is typically very weak so these techniques require high temporal resolution and high speed single-photon cameras with high detection efficiency.    \\
However, one of the main limitations of current NLOS imaging systems is the finite temporal resolution of the detectors, which in turn determines the spatial resolution of the retrieval and thus the ability to reconstruct satisfactorily the 3D structure of the hidden scene. Furthermore the complete 3D retrieval of the hidden scene often requires prohibitive time resources with the 3D imaging of a moving object, although progress is being made to reduce the acquisition and reconstruction times \cite{o2018confocal,goyal2019}, with the goal of reaching second or even sub-second times.  %Indeed  as the multiple back scattered signal is typically weak, this technique requires long acquisition times long \cite{velten2012recovering}, although recent results demonstrate 3D reconstruction at
%rates approaching 1 Hz for retroreflective hidden objects \cite{o2018confocal}. 
Additionally, the spatial resolution of the retrieval can be improved by using iterative backprojection algorithms that however, typically  require longer computational times \cite{la2018error,kak2001algorithms}.
\\
Another rapidly evolving imaging technique is based on so-called single-pixel cameras \cite{padgett2019}.  Standard 2D single-pixel imaging systems recover images by projecting an array of patterns onto the scene and detecting only the total reflected or transmitted light intensity, for which a single pixel is sufficient. The computational image reconstruction can then be achieved by computing a weighted sum of all the illumination patterns, where the weight of each pattern is given by the corresponding measured intensity level. This technique can also be extended for use e.g. in  microscopy \cite{radwell2014single}, imaging through scattering media \cite{tajahuerce2014image} and  terahertz imaging \cite{chan2008single} or full 3D LIDAR \cite{gibson2017real,sun2016single,edgar2016real}.\\
 Although this technique requires many consecutive measurements, the consequent long acquisition time required for high spatial resolution images can be significantly reduced by applying compressive sensing \cite{duarte2008single,satat2017lensless,chan2008single}. More importantly, the technique provides more flexibility in choosing the optimal (single pixel) detector for the imaging challenge being addressed.  Of relevance to this work, this implies that one can choose a single pixel detector with enhanced temporal response and build upon the typically better temporal resolutions that are available in single pixel format (timing resolutions down to picoseconds), when compared to camera/SPAD array technologies.\\
In this work we demonstrate full 3D retrieval of hidden scenes by using a time-resolved single-pixel camera. The choice of a single pixel camera approach allows us to use optimised (sub 30 ps impulse response function) single photon detectors in combination with a digital mirror device (DMD) so as to remove the need for any scanning components whilst building upon the 20 kHz refresh rate of the DMD and high single photon sensitivity to reduce acquisition times with good reconstruction fidelity.  By employing a white-light laser, we extend the technique to achieve the full RGB colour retrieval of non-cooperative, hidden objects and by choosing high efficiency single photon avalanche diodes (SPADs) we also achieve sub-second acquisition times. \\
\begin{figure}[t!]
	\centering
	\includegraphics[width=\linewidth]{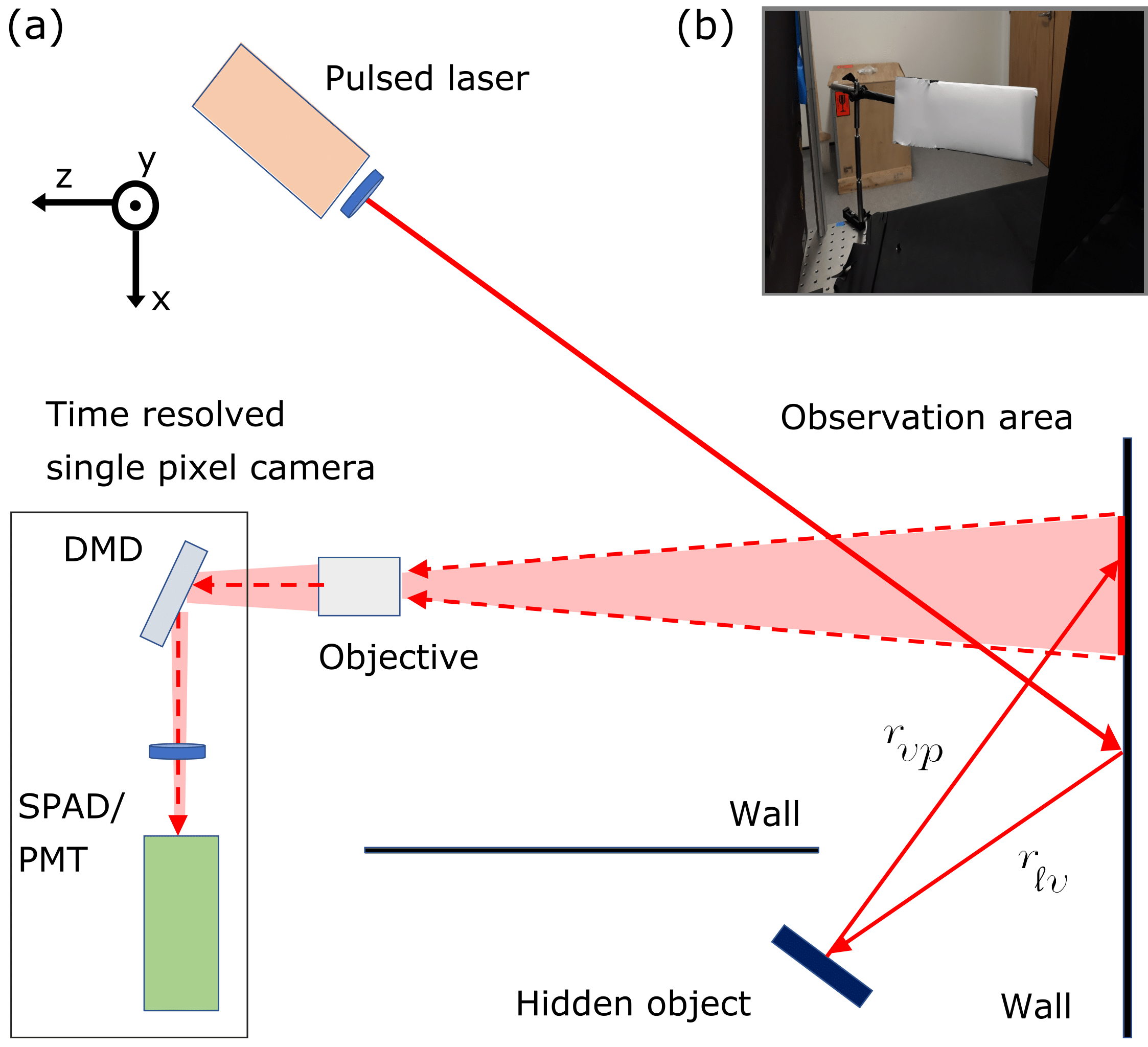}
	\caption{Schematics of the experimental set-up for non line of sight imaging. a) A pulsed laser scatters on a scattering surface producing a spherical wave in all the surrounding area. The DMD is imaging a 50x50 cm$^2$ area of the scattering surface and the time resolved single-pixel camera collects the signal back-scattered from the target in the hidden scene by temporal histograms. We collect the  signal scattered back by the hidden target by sequentially collecting light from each of the 20x20 pixels either by raster scan masks or by Hadamard masks. b) Picture of the non cooperative single object scene used for ultra-fast NLOS imaging.}
	\label{1}
\end{figure}
Figure~\ref{1}(a) shows the  experimental set-up: the single-pixel camera is composed of a camera lens objective (8 mm focal length, f/3.5) that images a 50x50 cm$^2$  portion of a scattering wall onto a DMD (placed 1.16 m from the wall). The DMD then projects only selected portions (masks) of the image onto a single-pixel single-photon detector through a microscope objective. The single-photon detector therefore acts as a bucket detector and has 30 ps impulse response time and an area of 57x57 $\mu$m$^2$. The DMD masks have 20x20 pixels corresponding to  2.6x2.6 cm$^2$ pixel areas on the scattering wall. The single-photon data is recorded in time-correlated-single-photon-counting (TCSPC) mode triggered by the illumination laser, as a histogram of photon arrival times with 4096 time bins of 6.1 ps duration each.  
The laser  is directed on the scattering wall 10 cm to the right of the field of view of the single-pixel camera, producing a first scattered spherical wave. Part of the spherical wave hits the hidden object which in turns scatters back into the field of view where it is captured by the time-resolved single-pixel camera. We then explored various imaging scenarios, testing different objects and DMD mask choices.\\
For the first scenario we investigated a hidden scene of two cooperative objects, i.e. a highly reflective, tin-foil 7.62 cm diameter cylinder and a 2.54 cm diameter mirror placed at different positions and distances from the wall  [see Fig.~\ref{2}(b)]. We used a 120 fs pulsed laser $808$ nm wavelength with a repetition rate of 80 MHz and an average power of 800 mW.  We collect the third bounce echo with a simple raster scan acquisition on the DMD, one  2.6x2.6 cm$^2$ imaging pixel at a time, and collect the reflected light onto a photomultiplier tube (PMT, hybrid photo detector HPM-100-07, Becker\&Hickl, 4\% efficiency at $808$  nm) with optimised temporal response (the measured total impulse response time was 27 ps FWHM). %The PMT has a round GaAsP hybrid detector with an active area of 3 mm of diameter and the light is focused on the sensor  with a 100 mm focal length lens. 
The acquisition  time is set by the amount of the back scattered signal detected by the sensor. In this case the acquisition time is 10 seconds per mask (i.e. pixel) for a total acquisition time of 66 minutes. % but data handling and  data processing increase the acquisition to 70 minutes. 
Figure~\ref{2}(a) shows one time frame of the collected third bounce echos of the  two hidden objects.\\ 
We proceed with the retrieval of the 3D shape of the hidden objects by applying the back-projection imaging algorithm first introduced by Velten et al. \cite{velten2012recovering,la2018error}. Although faster retrieval methods are available (see e.g. \cite{Arellano17}), our primary focus here is on the hardware rather than on the retrieval software. Regarding the time-of-flight evaluation, we consider the moment the laser hits the wall as the zero time reference. We therefore divide the 3D space in $10^6$ voxels and we calculate the likelihood of the target to be localized on each voxel by using % the time of flight information. %Thus, we  map the 3D space in a 3D matrix whose elements represent the voxels' likelihood.
%The likelihood of a voxel is based on 
the time of flight $t_p$ the light  takes to cover the distance $r_{\ell v}+r_{vp}$ where $r_{\ell v}$ is the distance between the laser and the voxel and $r_{vp}$ is the distance between the voxel and the pixel.  The relation $ct_p=r_{\ell v}+r_{vp}$ indicates that all the possible contributions to a given pixel lie on the surface of an ellipsoid whose foci are the laser spot and the pixel position. The ensemble of the 400 temporal histograms overlap at the scattering object position and thus encodes the 3D geometry of the hidden objects. %\textcolor{red}{YA: might be useful to add a figure showing the distances $r_{\ell v}$ and $r_{vp}$, and how $t_p$ is found?}
We assign the likelihood of a voxel by  summing the intensity of the pixels that could have received any contribution from the voxel.  %Each intensity is also  weighted by a  factor \textcolor{red}{XXX} to account for the distance attenuation of spherical waves and we take account of the distance between the pixels and the detector in the time of flight evaluation.   
%After obtaining the 3D  matrix of the voxels' reflectivity, we retrieve 
Following Ref.~\cite{velten2012recovering}, the final 3D shape of the object is improved by applying a Laplacian filter followed by a threshold selection on the data along the  $\hat{z}$ direction of the voxel grid. \\
Figures~\ref{2}(c) and (d) show the results on the x-y plane and on the x-z plane, respectively. %The image on the x-z plane is recovered by considering the  index along the $\hat{z}$ direction of the most likely voxel for each ($x-$$\bar{y}$) coordinate in Cartesian space where $\bar{y}$ is the average coordinate along  the $y$ direction. 
The dotted line in each figure indicates the  actual positions of the targets. Our results show that this technique provides an accurate 3D shape recovering of the hidden (cooperative) targets, although with relatively long acquisition times. \\
We then investigate a hidden scene of non-cooperative objects with the same setup (Figure~\ref{3}).  In this case, the scene to be recovered is a red-green-blue (RGB) coloured object, placed outside the direct line-of-sight [Fig.~\ref{3}(b)], where each coloured region has a rectangular shape of 20x9 cm$^2$. In order to retrieve color information we use a supercontinuum laser (SuperK EXTREME/FIANIUM, NKT Photonics, repetition rate 67 MHz, pulse duration $\sim10$ ps, average power 100 mW in the 450 nm - 700 nm spectral range).   For the RGB retrieval we run a separate measurement for each of the three RGB colours, using a corresponding band-pass spectral filter centred at of 490, 550 and 610 nm (40 nm bandwidth) after the laser source with roughly 20 mW average power for each color.  %We collect the third bounce echo by raster scan acquisition and collect the reflected light using  the same PMT detector with a  16\% efficiency at visible light. 
As above, the optimal acquisition time, due to the relatively low laser power, was found to be 10 seconds per mask for an overall acquisition time of 66 minutes. Figure~\ref{3}(a) shows the three return signals scattered back by the three  RGB coloured targets at a time frame of 8 ns. Figures~\ref{3}(c) and (d) show the reflectivity on the x-y plane and on the x-z plane respectively: as can be seen, the retrieved 3D scene corresponds very closely to the ground truth (dashed lines). \\% \textcolor{red}{YA: It might be worth mentioning whether the blue object appears in the red channel. Although primary blue, the paper might have a non-negligible reflectivity in the red and green channels?}\\
The last scenario we investigate is a non-cooperative hidden object (a white paper rectangle of 24x10 cm$^2$, see Fig.~\ref{1}(b)) aimed  at the optimising acquisition time. 
We achieve high speed acquisition by using a high efficiency (70 \% peak efficiency at  550 nm) SPAD detector \cite{sanzaro2018single} %manufactured in a 0.16 $\mu m$ BCD technology with monolithically integrated sensing circuit
 with a measured impulse response function of 30 ps FWHM. The SPAD has a square active area of 57x57 $\mu$m$^2$ and we use a 75 mm focal length lens and a long working distance objective (magnification factor 50) to focus the light after the DMD on to the sensor. The high sensitivity of the detector allows a shorter acquisition time of 1 millisecond per mask, which in this case where chosen as the first 400 Hadamard patterns with the goal of increasing the amount of collected light for each mask (50\% of the pixels are always projected onto the detector for each mask). For each Hadamard pattern, one binary mask and its negative are used and combined, leading to a total of 800 patterns. This allowed the total acquisition time to be reduced to only 0.8 s. In this case we used the same supercontinuum laser of the previous scenario with a power of 550 mW at 550 nm. %The backprojection algorithm has a run time of 47 seconds measured by using the Matlab tic-toc function.
\begin{figure}[t!]
	\centering
%	\subfloat[][\emph{}]
	{\includegraphics[width=\linewidth]{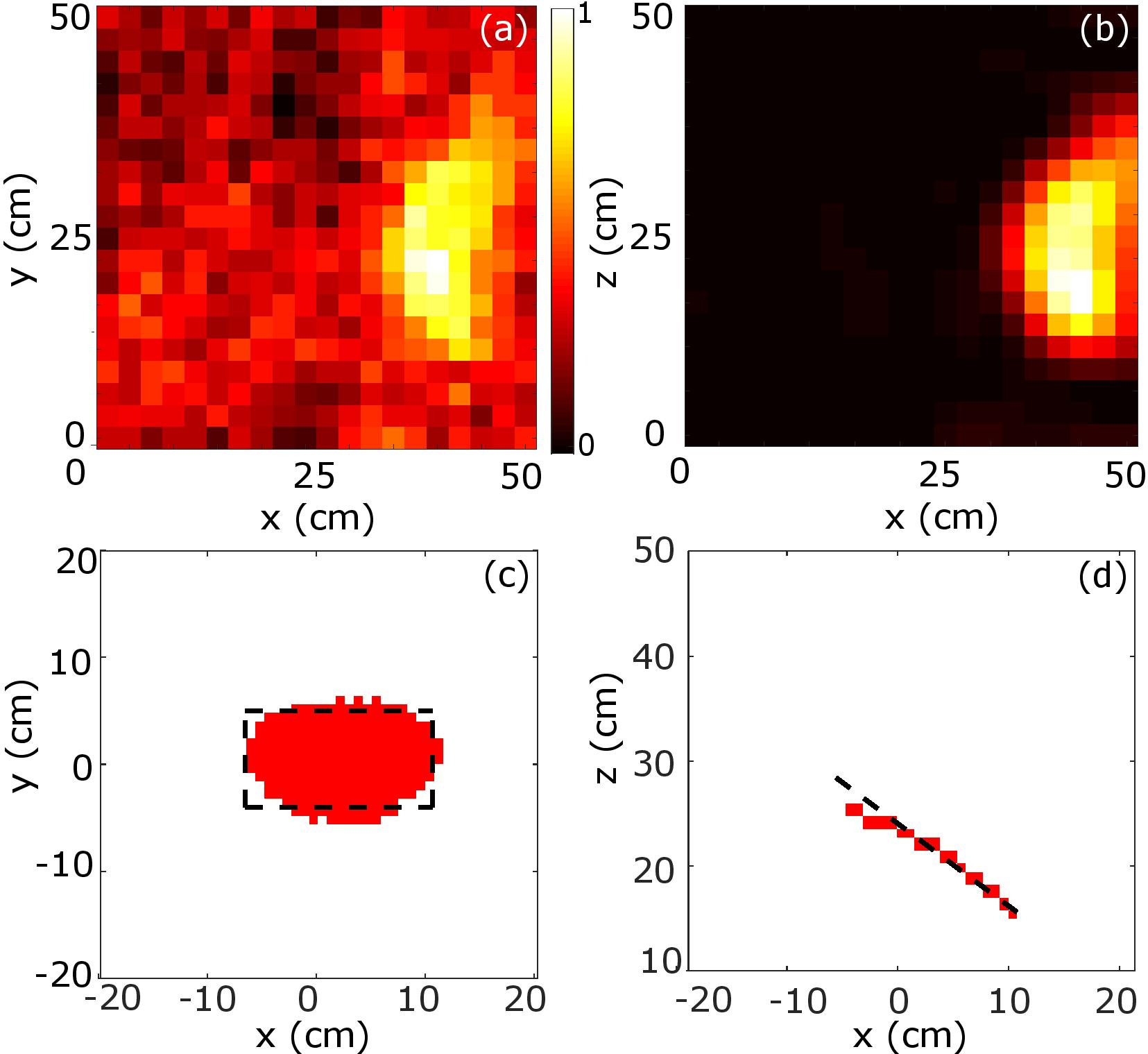}} \\
	\caption{Ultra-fast NLOS single-pixel camera results. a) Return signal  produced by the   hidden object by Hadamard acquisition at a time frame of 5.3 ns. b) Return signal obtained by pre-processing filter. c,d) Retrieval of the 3D shape of the hidden target in the x-y plane (c) and in the x-z plane applying backprojection imaging algorithm. The dotted black line represents the actual position of the target.  The object to be recovered is a rectangular target of 24x10 cm$^2$. The voxel dimension is 1.4x1.4x1 cm$^3$ for an overall of 140x140x100 cm$^3$. The threshold used for the retrieval is 0.80.}% e-f) Signal collected by a chosen pixel in time before and after the filtering process.}
	\label{4}
\end{figure}%
As shown in Fig.~\ref{4}(a) the collected third bounce echo of the signal is affected by low signal-to-noise ratio  due to the short acquisition time. %\textcolor{blue}{GM: as  the signal in time collected by a chosen pixel  (Fig.~\ref{4}(e))}. %\textcolor{red}{YA: Fig. 4(a) looks blurred and does not include the typical shot noise. In Fig.2 and 4, it might be useful to add a colorbar showing the light flux/photon counts}. 
We therefore first apply a denoising algorithm similar to that used in \cite{caramazza2018enhancing} thus returning the signal shown in Fig.~\ref{4}(b). Precisely, a cost function is defined, accounting for the forward model (including observation noise assumed to be Poisson distributed) relating the image sequence (or video) reaching the DMD and the set of temporal sequences recorded for each Hadamard pattern. The cost function also includes two penalty terms to promote temporal and spatial smoothness after denoising. This is enforced by using a spatial total variation (TV) regularization, as well as a low-pass constraint on the Fourier transform of the temporal intensity profile of each pixel. This cost function is convex and the denoising step, i.e. the cost function minimization, is performed using an alternating direction method of multipliers (ADMM) algorithm \cite{Figueiredo2010}, as in \cite{Altmann2016,Tobin2018}. %\textcolor{blue}{GM: Fig.~\ref{4}(f) shows the temporal signal of Fig.~\ref{4}(e) after the filtering process.}
Figures~\ref{4}(c) and (d) show  the retrieved reflectivity of the hidden objects on the x-y plane and on the x-z plane  respectively.  Our results therefore show that this technique  provides an accurate 3D shape of a hidden target even  with sub-seconds acquisition times { with an average number of only 1.2 photons/pixel in each time frame (with a maximum peak photon number of $\sim10$ photons/pixel)}. \\%\textcolor{red}{Can we include the average number of photons received for each pattern(or show raw data vs fitted temporal profile)? I suspect it low in this case and it would show how efficient the whole process is.}\\
In conclusion, the high efficiency and the high temporal resolution of single-pixel single-photon detectors allow us to accurately recover the 3D shape  of hidden objects even with low resolution masks of 20x20 pixels and no mechanical scanning parts.\\% as opposed to other systems requiring high number of pixels \cite{chen2018imaging} and many acquisitions \cite{velten2012recovering}.\\
The main limitation in the spatial resolution of the retrieval is due to the pixel size on the scattering wall. In our case, a pixel size of 2.6x2.6 cm$^2$ limits the temporal resolution to  60 ps due to the blurring of the pulse wavefront as it crosses the 3.6 cm pixel collection area (taking the diagonal of the square pixel). This effecively corresponds to an uncertainty in the arrival time of the return pulse and, in analogy with standard LIDAR, will translate into an uncertainty of the object depth location that is 1/2 this value, i.e. 1.8 cm. This can be overcome by decreasing the pixel size however, at the cost of longer acquisition times due to the larger number of Hadamard patterns (or pixels to scan).\\% Minor limitations to the spatial resolution of the retrieval are represented by the pulse length and the temporal resolution of the detector.\\
%The retrieval of scene hidden from the direct line of sight is a fundamental problem when the direct looking at the scene is impossible. 
Overall, the ability to identify a hidden scene by the proposed approach is mainly determined by the time resolution of the detector and by the time required to acquire a significant back-scattered signal. The retrieval of hidden scenes  still remains a challenging task due to long acquisition times, low spatial resolution of the retrieval and  computational sources, although significant steps have been made recently, see e.g. Refs.~\cite{o2018confocal,goyal2019}. We have shown a NLOS ultra-fast imaging technology which can reliably recover the 3D shape in colour of a scene with high spatial resolution by using single-pixel, single-photon detectors with high temporal resolution.   By using a high sensitivity  detector,  this system is able to retrieve the shape of the hidden target with  an improved acquisition time of 0.8 seconds paving the way to real-time 3D shape recovery of hidden objects. The accurate 3D shape recovery of this system could be further improved by fully exploiting the benefits of using single-pixel camera for the acquisition. Indeed a future improvement in this method would be  to decrease the acquisition times  by using  compressive sampling \cite{chen2018imaging}. By combining compressive sensing and constant improvements of  detection and computational resources, recovery of the 3D shape of hidden moving objects with high spatial resolution should be possible. \\
{\bf{Acknowledgments.}}
The authors acknowledge support from the Royal Academy of Engineering under the Research Fellowship scheme RF201617/16/31 and EPSRC, UK grant EP/M01326X/1.

\bibliographystyle{unstr}
%\bibliography{biblio}

\end{document}